\documentclass[prl,twocolumn,amsmath,amssymb,floatfix,showpacs,superscriptaddress]{revtex4}
\usepackage{graphicx}
\usepackage{amsmath}
\usepackage{bm}
\usepackage{epsf}

\newcommand{\bq}{\begin{equation}}
\newcommand{\ee}{\end{equation}}
\newcommand{\fr}[2]{\frac{#1}{#2}}
\newcommand{\eps}{\varepsilon}

\newcommand{\vM}{\vec M}

\newcommand{\vsigma}{\mbox{\boldmath $\sigma$}}

\newcommand{\hh}{h}
\newcommand{\epsgap}{\varepsilon_{\rm gap}}

\renewcommand{\vec}[1]{\mathbf{#1}}
\usepackage{colordvi}
\usepackage{graphicx}
\usepackage{color}
\newcommand{\red}{\color{red}}

\begin{document}
\title{Noiseless manipulation of helical edge state transport by a quantum magnet}
\date{\today }

\author
{P. G. Silvestrov} \affiliation{Institute for Mathematical
Physics, TU Braunschweig, 38106 Braunschweig, Germany}

\author{P. Recher}
\affiliation{Institute for Mathematical Physics, TU Braunschweig,
38106 Braunschweig, Germany} \affiliation{Laboratory for Emerging
Nanometrology Braunschweig, 38106 Braunschweig, Germany}

\author{ P. W. Brouwer}
\affiliation{Physics Department and Dahlem Center for Complex
Quantum Systems, Freie Universit\"{a}t Berlin, Arnimallee 14,
14195 Berlin, Germany}

\begin{abstract}
The current through a helical edge state of a quantum-spin-Hall
insulator may be fully transmitted through a magnetically gapped
region due to a combination of spin-transfer torque and spin
pumping [Meng {\em et al.}, Phys.\ Rev.\ B {\bf 90}, 205403
(2014)]. Using a scattering approach, we here argue that in such a
system the current is effectively carried by electrons with
energies below the magnet-induced gap and well below the Fermi
energy. This has striking consequences, such as the absence of
shot noise, an exponential suppression of thermal noise, and an
obstruction of thermal transport. For two helical edges covered by
the same quantum magnet, the device can act as robust noiseless
current splitter.
\end{abstract}
 \pacs{72.25.Pn,72.70.+m}
\maketitle

A time-reversal symmetry breaking magnetic field is well known to
introduce backscattering of helical edge states and to destroy the
conductance quantization of a quantum spin-Hall
insulator~\cite{KaneMele05,Bernevig06,HgTe/CdTe,InAs/GaSb}. More
subtle is the case of the effective magnetic field created by
dynamic spinful
impurities~\cite{Maciejko09,Tanaka11,Altshuler13,Yevtushenko15,Meng13}.
The reason for that is that a spin flip is necessary to reflect an
electron in a helical edge state. In the case of a small impurity
spin the magnetic impurities immediately become fully polarized,
leaving no room for any more backscattering of the
current~\cite{Maciejko09,Tanaka11}.

At first sight, the situation is different for a helical edge
coupled to a macroscopic magnet, because for a macroscopic magnet
the backscattering of a single electron in the helical edge
happens without a complete change of the magnet's polarization.
Moreover, the exchange coupling to a macroscopic magnet opens up a
gap in the helical-state spectrum, similar the gap opened by a
magnetic field. Yet, as was shown recently by Meng, Vishveshwara,
and Hughes \cite{Meng13}, under certain conditions concerning the
magnet's anisotropy energy, an electrical current incident on the
magnet is fully transmitted. Reference \cite{Meng13} invokes a
combination of spin-transfer torque and spin pumping
\cite{Slonczewski96,Berger96,Tserkovnyak2002,Tserkovnyak2005} as
the cause of this effect. The system considered in Ref.\
\cite{Meng13} was recently suggested as an ``adiabatic quantum
motor'' \cite{Arrachea15}.

In this letter we show that such a macroscopic magnet coupled to
the helical edge of a quantum spin-Hall insulator has very special
noise and thermal transport properties, some of which are
unparalleled in the field of mesoscopic quantum transport: Thermal
transport and shot noise are essentially absent in a two-terminal
setup, and a multiterminal geometry involving a magnet coupled to
two helical edges allows a current partitioning without shot
noise. The origin of these remarkable phenomena is that all
electrons close to the chemical potential $\mu$ are reflected if
$\mu$ is inside the magnet-induced gap, whereas the current is
effectively carried by electrons with energy below the gap, which
may be very large compared to both temperature and applied bias
--- a situation reminiscent of the inter-relation of electrons at
low and high energy in the case of the chiral anomaly
\cite{Nielsen1983}.

\begin{figure}
\epsfxsize=1.\hsize
\epsffile{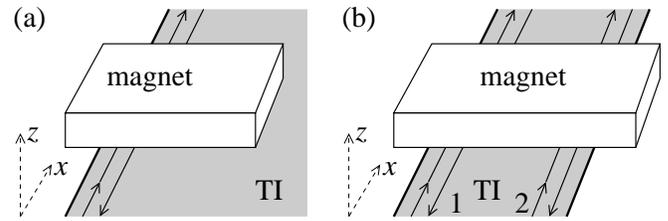} \caption{Schematic
drawing of the geometry we consider: One helical edge (a) or both
edges (b) of a two-dimensional topological insulator (grey)
exchange-coupled to a macroscopic magnetic
insulator.}\label{fig:1}
\end{figure}

We investigate the same system as in Ref.\ \cite{Meng13},
including the same conditions on the magnet's anisotropy energy
(see below), but we use the scattering
approach~\cite{Buettiker1992,BlanterBuettiker2000} to describe
transport. In addition to rederiving the results of Ref.\
\cite{Meng13} in a more general context, not being restricted to a
magnet that fully gaps the helical edge state, the scattering
approach gives us a unified framework for the description of
charge and energy transport and noise-related phenomena.
Reflection of helical electrons off the magnet is inelastic{\red,
} changing both their spin and energy when a bias voltage is
applied, the sign of the energy change depending on whether the
electron is incident from the source or the drain. Together,
transmitted electrons with energies below the magnet-induced gap
and reflected electrons in the helical edge form a noiseless
current-carrying state.

{\it Model and scattering matrix.--} The interaction of the
helical edge state with the magnet is described by the
second-quantized Hamiltonian \cite{Meng13}
 \begin{eqnarray}\label{Htot}
H = \int dx \hat{\boldsymbol\psi}^\dagger_{\bf x} \left[ -i\hbar
v_{\rm F}\partial_x\sigma_z + \hh(x) \vsigma \cdot \vM \right]
\hat{\boldsymbol\psi}_{\bf x}
 +\fr{D}{2} M_z^2.
 \end{eqnarray}
Here $v_{\rm F}$ is the Fermi velocity, $\sigma_{x,y,z}$ are the
Pauli matrices acting on the spinor $\hat{\boldsymbol\psi}_{\bf
x}=({\hat \psi}_{\uparrow}(x), {\hat \psi}_{\downarrow}(x))^{\rm
T}$ of the helical edge states and $\hh(x)$ is a function that
describes the exchange coupling between the magnetic moment $\vec
M$ and the edge state spin (both measured in units of $\hbar$),
such that $\hh(x)=0$ for $x \to \pm \infty$. A schematic picture
of the arrangement is shown in Fig.~\ref{fig:1}a.

We employ a macrospin approximation, {\em i.e.}, the magnetic
moment $\vec M$ in Eq.~(\ref{Htot}) is the only collective
variable describing the dynamics of the macroscopic magnet. The
last term in Eq.\ (\ref{Htot}) represents the magnetic anisotropy
energy. We take $D > 0$, corresponding to easy-plane anisotropy.
The Hamiltonian (\ref{Htot}) is invariant under spin-rotations in
the $x-y$ plane, so that the $z$ component of the total spin $M_z
+ \sigma_z$ is conserved. This additional symmetry of the model is
the key to the absence of backscattering in the steady state (in
the absence of residual interaction effects inside the edge)
\cite{Maciejko09,Tanaka11,Meng13}. We first derive this result for
the mean current, and then extend our discussion to noise and
distribution functions.

Prior to considering the full many body case it is convenient to
start with a single electron problem in the topological-insulator
edge interacting with the magnetic moment, described by the
first-quantized Hamiltonian
 \begin{equation}
H = -i\hbar v_{\rm F} \partial_x  \sigma_z + \hh(x) \vsigma \cdot
\vM + \frac{1}{2} D M_z^2.
  \label{Hsingle}
 \end{equation}
A reflection of an electron at the topological insulator edge is
accompanied by a unit change $M_z \to M_z \pm 1$ of the $z$
component of the magnetization, where the $+$ and $-$ signs refer
to electrons incident from the left or from the right,
respectively. The Hamiltonian (\ref{Hsingle}) thus decouples into
sectors in which the magnetization has the value $M_z$ for
right-moving electrons and $M_z+1$ for left-moving electrons.
Inside such a sector one may obtain a pure scattering problem by
performing the unitary transformation
 \begin{eqnarray}
  \tilde H = \begin{pmatrix} 1 & 0 \\ 0 & m_-\end{pmatrix} H
  \begin{pmatrix}1 & 0 \\ 0 & m_+\end{pmatrix},
 \end{eqnarray}
where $m_+ = (M_x+i M_y)/M_{\perp}$, with $M_{\perp} =
\sqrt{(M-M_z)(M+1+M_z)}${\red~,} is the operator that raises the
value of $M_z$ by unity. Using the equality $m_- m_+ = 1$ and
omitting constant terms, we have (cf.~\cite{Meng13})
 \begin{equation}
  \tilde H = \left[-i\hbar v_{\rm F}\partial_x + \hh(x) M_z -
\frac{\hbar \omega}{2} \right] \sigma_3
  + M_{\perp} \hh(x) \sigma_1   \ ,
  \label{eq:scat}
 \end{equation}
with $\hbar \omega = D(M_z + 1/2)$ the difference of the
anisotropy energies between the states with magnetization $M_z$
and $M_z + 1$. In Eq.~(\ref{eq:scat}) we neglected small terms
$\sim \hh$ in comparison to the large term $\sim \hh M_z$. In the
Hamiltonian $\tilde H$ the $z$ component of the magnetization can
be considered constant. The reflection and transmission amplitudes
$r(\varepsilon)$, $r'(\varepsilon)$, $t(\varepsilon)$, and
$t'(\varepsilon)$ for the scattering problem (\ref{eq:scat}) at
energy $\varepsilon$ can then be found using standard methods.
(Primed amplitudes are for electrons incident from the right.) In
particular for a smooth function $\hh(x)$ the Hamiltonian has a
gapped region with maximal gap $2 \epsgap$, with $\varepsilon_{\rm
gap} = \max_x h(x) M_{\perp}$. For our considerations it will be
important that $r(\varepsilon) \to 0$ if $|\varepsilon| \gg
\epsgap$.

The frequency $\omega$ appearing in Eq.~(\ref{eq:scat}) is the
frequency of rotation of the classical magnetization $\vec M$
around the $z$ axis. Note that outside the magnet region, the
kinetic energies of left-moving and the right-moving electrons are
$\eps_-=\varepsilon - \hbar \omega/2$ and $\eps_+=\varepsilon +
\hbar \omega/2$, respectively. Considering the kinetic energies
separately is important, because they appear in the distribution
function of incoming electrons, see Eq.~(\ref{fermid}) below.

Transforming back to the original formulation Eq.~(\ref{Hsingle}),
the scattering problem can be written in second-quantized form as
 \begin{align}
& \hat b_{\rm L}(\varepsilon_- ) = r(\varepsilon) m_+ \hat a_{\rm
L}(\varepsilon_+)
+ t'(\varepsilon) \hat a_{\rm R}(\varepsilon_-), \nonumber \\
& \hat b_{\rm R}(\varepsilon_+) = r'(\varepsilon) m_- \hat a_{\rm
R}(\varepsilon_-) + t(\varepsilon) \hat a_{\rm L}(\varepsilon_+),
  \label{eq:ba}
 \end{align}
where the operators $\hat b_{\rm L}(\varepsilon)$, $\hat b_{\rm
R}(\varepsilon)$ and $\hat a_{\rm L}(\varepsilon)$, $\hat a_{\rm
R}(\varepsilon)$ annihilate an {\em outgoing} and an {\em incoming}
electron~\cite{BlanterBuettiker2000} at a kinetic energy
$\varepsilon$, respectively, at the left (L) and the right (R) of
the magnet.

So far we have considered the problem of a single electron
scattering off the magnetic moment $\vM$. The scattering
amplitudes $r(\varepsilon)$, $r'(\varepsilon)$, $t(\varepsilon)$,
and $t'(\varepsilon)$, as well as the energy shift $\hbar \omega$
are functions of $M_z$. When considering the many-particle
problem, in principle, $M_z$ is a fluctuating quantity, because of
the simultaneous scattering off the magnetic moment of multiple
electrons. However, in the limit of a macroscopic magnetic moment
$M$ relative fluctuations of the out-of-plane magnetization $M_z$
are small and one may evaluate the amplitudes  $r(\varepsilon)$,
$r'(\varepsilon)$, $t(\varepsilon)$, and $t'(\varepsilon)$, as
well as the energy shift $\hbar \omega$ at the mean value $\langle
M_z \rangle$. With this approximation, Eq.\ (\ref{eq:ba}) can be
applied to the many-particle system.

{\it Current.--} The charge current through the helical edge is
calculated using the expression \cite{BlanterBuettiker2000}
 \begin{equation}
  I_{\rm L} = \frac{e}{h} \int d\varepsilon d\varepsilon'
  [a^{\dagger}_{\rm L}(\varepsilon) a_{\rm L}(\varepsilon')
  - b^{\dagger}_{\rm L}(\varepsilon) b_{\rm L}(\varepsilon')]
 \end{equation}
for the current to the left of the magnet, and a similar
expression for the current $I_{\rm R}$ to the right of the magnet.
For the incoming states one has
 \begin{equation}
 \label{fermid}
  \langle a^{\dagger}_{\alpha}(\varepsilon) a_{\beta}(\varepsilon) \rangle
= f_{\alpha}(\varepsilon) \delta(\varepsilon-\varepsilon')
\delta_{\alpha\beta},\ \ \alpha,\beta={\rm L},{\rm R},
 \end{equation}
where $f_{\alpha}(\varepsilon) = 1/[e^{(\varepsilon-
\mu_{\alpha})/k_{\rm B} T_{\alpha}}+1]$ is the distribution
function for reservoir $\alpha$, with chemical potential
$\mu_{\alpha}$ and temperature $T_{\alpha}$, $\alpha={\rm L}$,
${\rm R}$. Substituting Eq.\ (\ref{eq:ba}) and using
$|t(\varepsilon)|^2 = |t'(\varepsilon)|^2$, we find
 \bq\label{currentM}
  I_{\rm L} = \frac{e}{h} \int d\varepsilon [f_{\rm L}(\varepsilon_-) -
  |r(\varepsilon)|^2 f_{\rm L}(\varepsilon_+)
  -
  |t(\varepsilon)|^2 f_{\rm R}(\varepsilon_-)].
 \ee

When a bias voltage $eV$ is applied across the magnet,
 \bq
  \mu_{\rm L} = \mu + e V/2,\ \ \mu_{\rm R} = \mu - eV/2,
 \ee
initially the reflections of electrons incident from the left and
from the right will not be in balance. Since each reflection leads
to a change $\Delta M_z = \pm 1$, with a $+$ sign for electrons
incident from the left and a $-$ sign for electrons incident from
the right, the application of a bias leads to a finite
out-of-plane magnetization component $M_z$. The rate of change of
$M_z$ is the difference of reflections rates for electrons
incident from the left and from the right,
 \begin{equation}
  \langle \dot{M_z} \rangle = \fr{1}{h}
  \int d\varepsilon |r(\varepsilon)|^2 [f_{\rm L}(\varepsilon_+)
  - f_{\rm R}(\varepsilon_-)].
 \label{stationary}
 \end{equation}
Note that the integral is convergent, because $r(\varepsilon) \to
0$ for $|\varepsilon| \gg \epsgap$. If the two contacts are held
at the same temperature, the integrand in Eq.~(\ref{stationary})
never changes sign, and the stationary condition $\langle
\dot{M_z} \rangle = 0$ may be achieved at $\hbar\omega =eV$ only,
{\it i.e.}, when the integrand vanishes at all $\eps$.
Substitution of Eq.~(\ref{stationary}) into (\ref{currentM}) then
yields the stationary current (using $|r(\varepsilon)|^2 +
|t(\varepsilon)|^2 = 1$)
 \begin{equation}
  I_{\rm R} = I_{\rm L}=\fr{e}{h}
  \int d\varepsilon [f_{\rm L}(\varepsilon_-)
  - f_{\rm R}(\varepsilon_-)] = \frac{e^2}{h} V.
  \label{current}
 \end{equation}
This is the reflectionless current originally obtained by Meng
{\em et al.} \cite{Meng13}.

\begin{figure}
\vspace{1.cm}
\includegraphics[width=8.5cm]{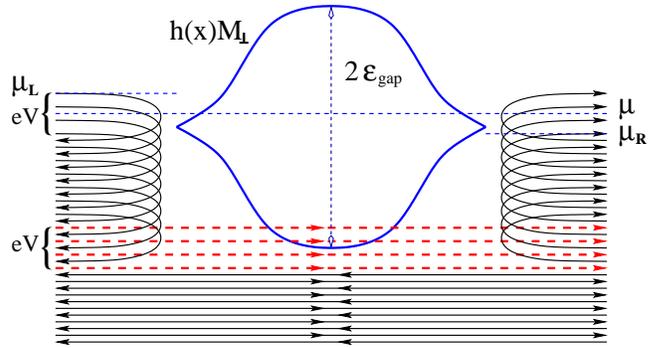}
\caption{Schematic illustration of reflection and transmission as
a function of energy, for the special case that the exchange
coupling $\hh(x)$ is a smooth function of $x$ and the chemical
potential $\mu$ lies inside the spectral gap induced by the
coupling to the ferromagnet. The vertical axis refers to the
electrons' kinetic energy, which changes by an amount
$\hbar\omega =eV$ upon reflection. Electrons carrying the actual
current are shown red/dashed. Although details change if the
coupling function $\hh(x)$ is not smooth, the conclusion that the
current is carried effectively by electrons far away from the
Fermi level remains true as long as the chemical potential is
inside the gap. } \label{fig.current}
\end{figure}

In Fig.\ \ref{fig.current} we schematically illustrate which
electrons carry the current in the special case that the function
$\hh(x)$ is smooth, so that the reflection probability
$|r(\varepsilon)|^2 = \Theta(\epsgap-|\varepsilon|)$, $\Theta$
being the step function, and with chemical potential $\mu$ in the
gap. Since the kinetic energy for electrons incoming from the left
reservoir is $\varepsilon_{+} = \varepsilon + \hbar \omega/2$,
electrons incident from the left are fully reflected if (and only
if) their {\em kinetic} energy $-\epsgap + \hbar \omega/2 <
\varepsilon_{+} < \epsgap + \hbar \omega/2$. Similarly, electrons
incident from the right are fully reflected if and only if their
kinetic energy $\varepsilon_{-}$ is between $-\epsgap - \hbar
\omega/2$ and $\epsgap - \hbar \omega/2$. Hence, the current is
carried effectively by electrons incident from the left, with
kinetic energies between $-\epsgap - \hbar \omega/2$ and $-\epsgap
+ \hbar \omega/2$. These are electrons far below the Fermi level,
as shown in the figure.

Note that, while this qualitative picture relies on the chemical
potential $\mu$ being inside the gap, the conclusion
(\ref{current}) of a perfectly transmitted current does not rely
on this condition. The only change in the case of a chemical
potential outside the gap is an increased relaxation time because
of the small reflection coefficient $r(\eps)$ for energies near
$\mu$, see Eq.~(\ref{stationary}).

{\it Noise.--} Since Eq.~(\ref{current}) predicts a perfect
transmission of the current, one should expect no zero-frequency
shot noise. This follows directly for the special case considered
in Fig.~\ref{fig.current}, where all right moving states with
energies below $\mu_{\rm L}$ and all the left moving states with
energies below $\mu_{\rm R}$ are occupied, leaving no room for any
uncertainty, {\em i.e.}, for noise. Figure \ref{fig.current} also
suggests a strong suppression of the thermal noise in the case of
the chemical potential inside the gap, since in that case all
electrons with energy near the chemical potential --- {\em i.e.},
all electrons that ``know'' about the temperature --- are
reflected.

To formally calculate the noise we may use the scattering matrix
(\ref{eq:ba}). The zero-frequency noise power then takes the form
(cf.\ Ref.\ \cite{BlanterBuettiker2000})
 \begin{multline}
S=\frac{2e^2}{h}\int d\varepsilon \,
  \left\{
|t(\varepsilon)|^2[f_{\rm L}(\varepsilon_+)\left(1-f_{\rm
L}(\varepsilon_+)\right)^{\vphantom{33}}\right.\\
\mbox{} + f_{\rm R}(\varepsilon_-)(1-f_{\rm R}(\varepsilon_-))]\\
  \left. \mbox{}
  + |t(\varepsilon)|^2\left(1-|t(\varepsilon)|^2\right)\left(f_{\rm
L}(\varepsilon_+)-f_{\rm R}(\varepsilon_-)\right)^2 \right\} .
\label{eqnoise1}
 \end{multline}
For equal lead temperatures $T = T_{\rm R} = T_{\rm L}$ and in the
stationary limit $\hbar\omega=eV$ one has $f_{\rm
R}(\varepsilon_-)= f_{\rm L}(\varepsilon_+)$, see
Eq.~(\ref{stationary}), so that the last line in Eq.\
(\ref{eqnoise1}), the shot noise, vanishes. What remains is the
thermal noise, which, upon using the equality $f_{\rm
R}(\varepsilon_-)= f_{\rm L}(\varepsilon_+)$, becomes
 \begin{equation}
  S=\frac{4e^2}{h}\int d\varepsilon\,|t(\varepsilon)|^2 f_{\rm L}
(\varepsilon_+)\left(1-f_{\rm L}(\varepsilon_+)\right).
 \label{eqnoise2}
 \end{equation}
The thermal noise depends in general on the specific form of
$t(\varepsilon)$. For a smooth $\hh(x)$ one has
$|t(\varepsilon)|^2 = \Theta(|\varepsilon|-\epsgap)$, the energy
integration is easily done and one finds
 \begin{equation}\label{noisesedge}
  S=\frac{8e^2 k_{\rm B} T}{h}
  e^{-\epsgap/k_{\rm B} T} \cosh(\mu/ k_{\rm B} T),
 \end{equation}
plus corrections that vanish in the limit $|\epsgap \pm \mu| \gg
k_{\rm B} T$. If the function $\hh(x)$ is not smooth the detailed
expression for the shot noise power changes, but not the
conclusion that $S$ is exponentially small in $\min(\epsgap \pm
\mu)$. Clearly the exponential suppression of the thermal noise
indicates a departure from the usual form $S_{\rm Th}= 4k_BT
dI/dV$ for two-terminal conductors \cite{BlanterBuettiker2000}.
For chemical potential $\mu$ well outside the gap region the
thermal noise obtained from Eq.\ (\ref{eqnoise2}) agrees with the
conventional result $S = S_{\rm Th}$.

\begin{figure}
\vspace{1.cm}
\includegraphics[width=8.5cm]{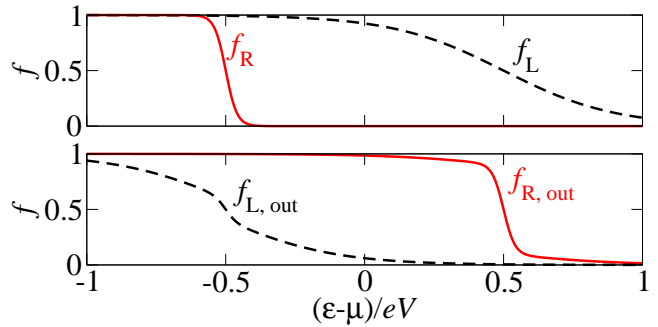}
\caption{Illustration of distribution functions for incoming (top)
and outgoing (bottom) electrons, for the case $T_{\rm L}/eV =
0.2$, $T_{\rm R}/eV = 0.02$, $|r(\varepsilon)|^2 = 1 -
|t(\varepsilon)|^2 = 0.8$. The location of the ``step'' in the
distribution function for the outgoing electrons corresponds to
that of the opposite reservoir, consistent with the perfect
transmission of the incident current. For this example, the width
of the step is, however, predominantly that of the same-side
reservoir, consistent with the fact that the device acts as a
``thermal insulator''.
 }\label{fig.two_temperatures}
\end{figure}

{\it Distribution functions.--} The strong suppression of thermal
noise can also be illustrated through a calculation of the
distribution functions $f_{{\rm R, out}}$ and $f_{{\rm L, out}}$
for electrons, which have been reflected from or transmitted
through the device. One finds
 \begin{align}\label{distribution}
f_{{\rm R, out}}(\eps)=f_{\rm L}(\eps)|t(\eps_-)|^2 +f_{\rm
R}(\eps-\hbar\omega)|r(\eps_-)|^2 ,\nonumber\\
f_{{\rm L, out}}(\eps)=f_{\rm L}(\eps+\hbar\omega)|r(\eps_+)|^2
+f_{\rm R}(\eps)|t(\eps_+)|^2,
 \end{align}
where $f_{\rm R}$ and $f_{\rm L}$ are the distribution functions
of electrons incident from the right and left reservoirs,
respectively. For equal reservoir temperatures and in the
stationary limit $\hbar\omega =eV$, such that $f_{\rm
L}(\eps)=f_{\rm R}(\eps-\hbar\omega)$, this equations reduce to
$f_{{\rm R,\, out}}(\eps)= f_{\rm L}(\eps)$ and $f_{{\rm L,\,
out}}(\eps)= f_{\rm R}(\eps)$ in accordance with
Eq.~(\ref{current}). However, the distributions become nontrivial
if the temperatures $T_{\rm L}$ and $T_{\rm R}$ are different. If
the chemical potential $\mu$ is inside the gap, all electrons with
energy close to $\mu$ are reflected, and one finds $f_{{\rm R,\,
out}}(\eps) = f_{{\rm R}}(\eps - \hbar \omega)$, $f_{{\rm L,\,
out}}(\eps) = f_{{\rm L}}(\eps - \hbar \omega)$, {\em i.e.}, the
device transmits the charge of the incident electrons, but not
their ``temperature'' --- the device is a thermal insulator. An
example for generic reflection and transmission amplitudes is
shown in Fig.~\ref{fig.two_temperatures}

{\it Noiseless partitioning of the current.--} It is instructive
to also consider a four-terminal quantum spin Hall device with two
helical edges covered by the same magnet, as in Fig.~\ref{fig:1}b.
Let the lower left contact be biased by a voltage $V$ and all
other terminals be put to ground. The electrons incident from the
lower left contact initiate a precessing non-equilibrium
out-of-plane magnetization $M_z$, which in turn drives spin and
charge currents in the remaining three terminals. The scattering
approach used above can be carried over straightforwardly for each
edge, as there is no scattering between them. The precession
frequency $\omega$ at which a steady state sets in now is given by
 \begin{align}
0=\langle &{\dot M_z}\rangle=\fr{1}{h}\int d\varepsilon
[|r_1(\varepsilon)|^2\left(f_{\rm 1L}(\varepsilon_+)-f_{\rm
1R}(\varepsilon_-)\right)\nonumber\\
&\left.-|r_2(\varepsilon)|^2(f_{\rm 2L}(\varepsilon_-)-f_{\rm
2R}(\varepsilon_+)\right)] \ , \label{stationary2}
 \end{align}
where the labels $1$ and $2$ refer to the two edges and we have
used that the helicity of the edge states is opposite in the two
edges. The ability of the magnet to create/change current in each
edge state is determined by the reflection coefficient leading to
different currents $I_1$ and $I_2$ for arbitrary
$r_{1,2}(\varepsilon)$. However, in the case of all chemical
potentials inside the magnet-induced gap with exponential accuracy
one has $|r_1(\varepsilon)| = |r_2(\varepsilon)|=1$, leading to
$I_1 = -I_2 = e^2 V/2h$, independent (again with exponential
accuracy) of the different temperatures of the four contacts.

To calculate the noise power one may use Eq.~(\ref{eqnoise1}) for
each edge separately. In the example that all chemical potentials
are inside the gap, the currents are obviously noiseless, up to
exponentially small corrections in $|\epsgap \pm \mu|/k_{\rm B}T$
--- a result that can already be understood by arguing that in
both helical edges the current is carried by electrons far below
the chemical potential, in a picture very similar to that of Fig.\
\ref{fig.current}. The absence of noise may be considered
surprising, since, unlike in the two-terminal setup, in the
four-terminal setup the original incident current appears to be
partitioned.

{\it Conclusions.--} We considered the transport of helical edge
state electrons in the proximity to a magnet with easy-plane
anisotropy compatible with the spin helicity of the edge state.
While it was known that (after transient effects) such a system
perfectly transmits an incident charge current \cite{Meng13}, in
spite of the fact that the coupling to the magnetic insulator
opens a gap in the spectrum of the helical edge, we have shown
that the device has very special noise properties: The current is
noiseless, and thermal transport is blocked. We explain this
combination of ``perfect metal'' and ``perfect thermal insulator''
properties in a single-particle scattering picture, in which
effectively the current is carried by electrons with energy far
below the chemical potential. In a four-terminal setup, the same
device can be used as a noiseless current splitting device.

Our predictions rely strongly on the precise orientation of the
magnet's easy-plane anisotropy. While the orientation chosen here
is generic for a thin magnetic film~\cite{oHandley2000}
exchange-coupled to the spin polarized helical edge
modes~\cite{bruene2012} of a quantum spin-Hall material,
small deviations from the ideal limit may still exist. These will
be investigated in future work.

We acknowledge discussions with B. Probst. This work was supported
by the DFG grant RE~2978/1-1 and by the SFB 658 ``molecular
switches on surfaces''.

\end{document}